# Non Linear Software Documentation with Interactive Code Examples

MATHIEU NASSIF and MARTIN P. ROBILLARD, McGill University, Canada

Documentation enables sharing knowledge between the developers of a technology and its users. Creating quality documents, however, is challenging: Documents must satisfy the needs of a large audience without being overwhelming for individuals. We address this challenge with a new document format, named Casdoc. Casdoc documents are interactive resources centered around code examples for programmers. Explanations of the code elements are presented as annotations that the readers reveal based on their needs. We evaluated Casdoc in a field study with over 300 participants who used 126 documents as part of a software design course. The majority of participants adopted Casdoc instead of a baseline format during the study. We observed that interactive documents can contain more information than static documents without being distracting to readers. We also gathered insights into five aspects of Casdoc that can be applied to other formats, and propose five guidelines to improve navigability in online documents.

## 1 INTRODUCTION

We present Casdoc, a novel technology for improving the presentation of online learning resources for programmers. Casdoc, which stands for Cascading documentation, presents the content of an HTML document as a graph of concise and interactive annotations rooted in a code example. A transformation tool simplifies the authoring process of these documents by generating them from annotated code files.

Novel presentation approaches are needed to improve the way information seekers, such as programmers, use documentation. Documentation is a crucial asset to understand an unfamiliar software system [23, 58]. Yet, creating good documents requires a lot of effort and expertise. Software engineering researchers have proposed different techniques to generate content (e.g., [11, 39, 50]) or retrieve it from knowledge bases (e.g., [17, 55, 75]), which can alleviate some of this effort. However, documentation quality is multi-faceted [32]: it must not only contain enough information to address the concrete needs of its audience [15], but the information must also be readable, navigable, and understandable [2, 74]. These aspects, which relate to *how* the information is organized and presented, have not been studied as extensively. As a result, current documentation formats may fail to emphasize the most useful fragments when too much content is available [77].

Casdoc is a solution to improve the navigability of content in code-oriented documents. In a Casdoc document, readers interact with code elements to reveal further explanations of those elements (see Figure 1 in Section 2). Information about elements that are irrelevant to a reader remains hidden to avoid unnecessary distractions. Casdoc relies on popovers and dialogs to achieve this objective. Hence, it recasts two graphical elements which are typically used for secondary navigation aid as the primary structure to organize the content of a document. As a result, this strategy splits the content of a document into concise annotations. Annotations are created by the document's author, who inserts them directly into working code files as code comments. This authoring process is similar to the generation of API reference documentation from header comments, but supports a different type of documentation (i.e., tutorials and other learning-oriented documents). By writing each explanation in isolation, authors do not need to concern

Authors' address: Mathieu Nassif, mnassif@cs.mcgill.ca; Martin P. Robillard, robillard@acm.org, School of Computer Science, McGill University, Montréal, QC, Canada.

themselves with the narrative flow of the document. The Casdoc transformation tool then converts the annotated code files into dynamic web documents.

Previous work has suggested approaches to improve the format of learning resources, as the transition from printed to digital documents created opportunities for new modes of interaction [29, 71]. Researchers have proposed techniques to add interactive elements to existing static documents, such as data visualizations [5, 45, 46], custom annotations [31], and automatically generated links to external resources [4]. Other work has suggested to make document visualization software more interactive with navigation features inspired by paper-based formats [63, 65]. Specifically for software documentation, researchers proposed to augment the interface of code search tools with explanations of how the code snippets were matched, to help programmers decide on the pertinence of the results [73]. Digital documents can also contain dynamic elements, such as runnable code examples [47, 72], explorable statistical analyses [18], or modifiable machine learning models [6]. All of these techniques, however, do not change the linear organization of information in existing documents, which does not always match the readers' navigation patterns [10, 30].

Casdoc challenges this traditional structure. We observed how programmers react to a non linear format in a seven-month field study with 326 participants and 126 documents. Participants were undergraduate students enrolled in a programming-intensive software design course, who used the documents to learn professional software design know-how. We designed the field study to maximize its ecological validity and avoid interference with participants, as they should prioritize the course's learning objective over their participation in the study. We analyzed over 18 000 participant actions on the documents to assess the strengths and limitations of five presentation aspects of Casdoc that can be replicated in other presentation formats. Based on our results, we propose guidelines for designing new code-oriented documentation formats and improving existing ones. We also leveraged these findings and feedback received on preliminary versions of this work [51, 52] to improve Casdoc. We released a public set of learning resources for software design that use the improved version of Casdoc. This article makes the following contributions:

(1) the description of an interactive and non linear format for software documents, which addresses common documentation issues (Section 2 and 5.3);

(2) an analysis of five relevant document design factors, based on prior work (Section 3);

(3) a complete methodology for the design of a field study that maximizes the ecological validity and reliability of the results in a context where the investigators have authority over participants (Section 4);

(4) the results of our study, synthesized into guidelines for designing new software-oriented documentation formats (Section 5).

*Replication.* The material necessary to replicate our study is publicly available from the Casdoc project's web page, at https://www.cs.mcgill.ca/~martin/casdoc/. Readers can find on this page a free online service to convert annotated code files into the preliminary version of Casdoc used during the field study, with a detailed description of the annotation syntax. The page also includes links to the course textbook, which has a public companion website, and to the annotated code examples used in the study. The annotated code examples have been updated to the newest version of Casdoc, but their content is similar to what was available to participants during the field study. We do not publicly release the database of interaction events collected during the field study to protect the participants' information.

## 2   THE CASDOC DOCUMENTATION FORMAT

Casdoc is a *presentation format* for online programming learning resources. Casdoc documents present a central code example, with additional explanations as interactive annotations. Authors create documents by writing regular source

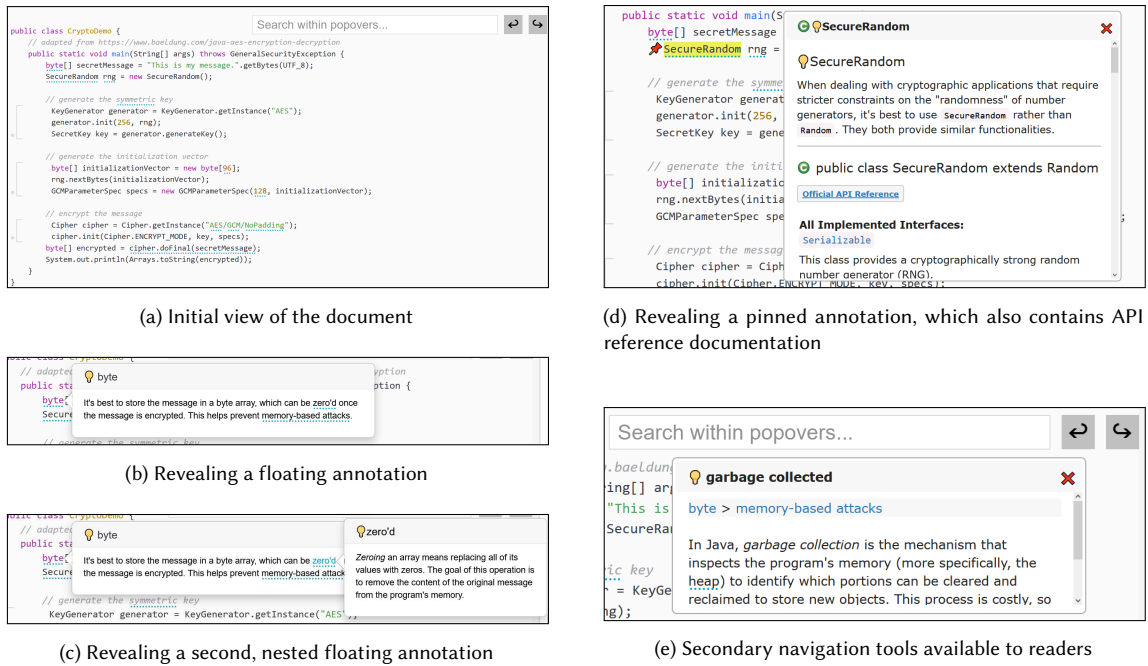

(a) Initial view of the document

(b) Revealing a floating annotation

(c) Revealing a second, nested floating annotation

(d) Revealing a pinned annotation, which also contains API reference documentation

(e) Secondary navigation tools available to readers

Fig. 1. Example of a Casdoc Document with Different Annotations Revealed

code files and inserting explanations in-place as code comments. The Casdoc *transformation tool* then converts the annotated code files into interactive web documents. Our implementation currently supports code examples written in the Java programming language.

Casdoc is designed for learning resources that focus on the implementation of programming concepts, such as programming forum posts and tutorials. It can demonstrate how to use a programming technology or the realization of programming concepts such as design patterns. In contrast, Casdoc is not intended for internal developer documentation and documents that focus on theoretical concepts.

## 2.1 Presentation Format

Figure 1 presents five views of a Casdoc document.[1] The initial view of the document shows only a central code example, which acts as the *root* of the document. For example, Figure 1a shows a code example that illustrates how to use Java's cryptography application programming interface (API) to encrypt a message.

Additional explanations of the root code example are placed in *annotations*. Annotations are interactive elements that are overlaid on top of the code example. They are hidden in the initial view of the document. Readers can selectively reveal the annotations that contain information relevant to them, then hide them again once they no longer need the information.

---

[1]The details of the format shown in Figure 1 is consistent with a preliminary version used in field testing. We introduced it first as a tool demonstration at a conference [51]. We made several improvements to the format, including visual modifications, based on the results of the study. Those improvements are described in Section 5.3. Nevertheless, the description of Casdoc in this section applies to both versions of the format.

Each annotation contains some information about a specific code element, called its *anchor*. Anchors have visual *markers*, which indicate the presence of additional explanations to the reader. The anchor of an annotation can be any string of text on a single line (*inline anchor*) or any continuous set of lines (*block anchor*) in the code example. Figure 1b shows an annotation, anchored on the keyword byte, that explains why the original message is stored in a byte array as opposed to a String object. Some annotations can be associated with multiple anchors, for example when an important code elements appears multiple times.

The anchor of an annotation can also be a string of text *inside* another annotation, such as the mention of an important concept. In this case, the annotation that contains the anchor is the *parent* annotation, and the annotation that the anchor links to is a *nested* annotation. Figure 1c shows a nested annotation that defines the expression "zeroing an array", which appears in the annotation about the byte array. Nested annotations can themselves contain other nested annotations.

Readers can view annotations in two forms. Hovering over an anchor reveals a *floating* annotation, which disappears when the reader leaves the area of the anchor and its annotation (Figure 1b). Clicking on the anchor *pins* the annotation, keeping it visible until the reader clicks again on the anchor (Figure 1d). Readers can move and resize pinned annotations.

Typical annotations come from the comments inserted by the document's author in the annotated code file. However, the Casdoc transformation tool automatically creates additional annotations with the official API reference documentation for standard Java types and their members (*Javadoc* annotations), anchored on the type or member's name in the code.[2] By contrast, annotations created by the document's author are referred to as *original* annotations. If the anchor of a Javadoc annotation overlaps with the anchor of an original annotation, the two annotations are combined into a single one, with the two fragments clearly separated. Figure 1d shows an annotation that contains both the rationale for using the SecureRandom class instead of the more usual Random to generate numbers, and the reference documentation of that class.

To help readers orient themselves across the graph of annotations, Casdoc includes several visual aids and navigation tools. When an annotation is pinned, a pin icon appears beside its anchor, and the anchor is highlighted when the reader hovers over the annotation (Figure 1d). Pinned nested annotations show a *breadcrumb trail* to indicate their parents and allow readers to open them (Figure 1e). Readers can also use a custom *search bar* to search among the content of all annotations.[3] Finally, readers can *undo* and *redo* pinning and unpinning actions, e.g., in case they accidentally close a nested annotation and forgot where the anchor was.

## 2.2   Authoring Process

To create a Casdoc document, an author starts by providing the root code example in a new Java file. The author then inserts annotations in code comments next to their anchors. Embedding annotations in a code file provides a format familiar to programmers. It ensures that the root document can be created and maintained using common development tools, such as integrated development environments (IDEs) and version control systems (e.g., Git). Each Java file will be converted into a separate Casdoc document.

Authors use a special syntax to distinguish code comments that contain annotations from regular comments to keep in the final document. Each annotation comment is enclosed between the sequences /*? and */, and may contain multiple annotations. Within a comment, each annotation starts with the declaration of its anchor, followed by its content.

---

[2]The anchor of Javadoc annotations are not indicated by markers, to avoid too many anchors and because the presence of the annotation is predictable.
[3]The native search feature of web browsers cannot reveal or find text in hidden annotations.

Authors can use the Markdown syntax to create visually rich annotations [24]. After inserting the desired annotations, the author uses the transformation tool to convert the annotated Java file into the final Casdoc web document.

The special syntax for declaring annotations does not interfere with the original Java code. Therefore, the annotated files can be validated for syntax and symbol resolution by any compatible compiler

### 2.3 Implementation

Casdoc documents are self-contained. The HTML file generated by the transformation tool contains all declared annotations, using dedicated HTML elements to identify them and their anchors. The visual elements and interactive aspects of the format are implemented with client-side CSS and JavaScript assets, which themselves only rely on mature libraries.[4] Hence, Casdoc documents can be viewed in any software that supports standard web technologies and can be deployed easily without requiring a complex server infrastructure.

The transformation tool is implemented as a Java program. It relies on a Java-specific parsing and symbol resolution library to extract custom annotations from code comments and Javadoc annotations related to standard types and members. A preliminary version of the tool is available as a free online service from the project web page.[5]

## 3 KEY PROPERTIES OF CASDOC

Documentation formats can vary across numerous dimensions, such as the length of code examples, the interplay between text, figures, and code, or the use of external resources as integral or peripheral information sources [3, 26]. The creation of Casdoc involved many decisions, from core design principles to technical implementation details. Not all of those decisions, however, have the same impact on the ways readers find information in documents.

We identified five properties of Casdoc that are key components of the format. Casdoc documents

(1) focus on code examples,
(2) gradually reveal information,
(3) split information into small fragments,
(4) use explicit hints about the information structure, and
(5) integrate content from external sources.

The properties are informed by prior work on programmer information needs and reading behavior. Each property thus corresponds to a hypothesis, namely that the property will help readers locate the information they need within a document. We used these properties to scope our evaluation of Casdoc. This scope focuses our findings on aspects of documentation that can be found in other existing formats or that can be used to design new ones.

We present each property with the prior work that supports it, its realization in Casdoc, and other examples of the property found in existing documentation. We also discuss potential limitations of the property, or contexts in which it may be detrimental to a document's quality. Table 1 shows an overview of the properties across a sample of documentation sources and formats. Those examples demonstrate alternative implementations of the properties. We note that the presence or absence of a property does not correlate with the overall quality or usefulness of the documents. In particular, the official Java tutorials (second row), despite being high-quality documents, do not exhibit any of the studied properties. We selected all resources as examples of good documentation. The objective of this work is to explore various strategies for presenting information, rather than try to rank different formats.

---

[4]The web assets can optionally be embedded in the Casdoc document, to make it a truly independent HTML file.
[5]https://www.cs.mcgill.ca/~martin/casdoc/

Table 1. Presence of the Five Properties in Documents from Various Sources

| Document Source<br>Link | Focus<br>on Code | Gradual<br>Reveal | Small<br>Fragments | Explicit<br>Hints | External<br>Content |
|---|---|---|---|---|---|
| Casdoc<br>https://www.cs.mcgill.ca/~martin/casdoc/ | ✓ | ✓ | ✓ | ✓ | ✓ |
| Oracle's Java Tutorials<br>https://docs.oracle.com/javase/tutorial/java/index.html | - | - | - | - | - |
| Java API documentation (Javadoc)<br>https://docs.oracle.com/en/java/javase/17/docs/api/ | - | - | ✓ | ✓ | - |
| Stack Overflow<br>https://stackoverflow.com/questions | - | - | ✓ | - | - |
| Android Developer Guides<br>https://developer.android.com/topic/architecture | - | ✓ | - | ✓ | - |
| Amazon API Gateway's FAQs<br>https://aws.amazon.com/api-gateway/faqs/ | - | ✓ | ✓ | - | - |
| R Cookbook<br>https://rc2e.com/ | ✓ | - | ✓ | - | - |
| Codelets [53]<br>https://dl.acm.org/doi/10.1145/2207676.2208664 | ✓ | ✓ | ✓ | ✓ | - |
| Adamite [31]<br>https://adamite.netlify.app/ | - | ✓ | ✓ | ✓ | - |
| SISE [67]<br>https://dl.acm.org/doi/10.1145/2884781.2884800 | - | ✓ | ✓ | - | ✓ |

We discuss these properties as they apply within the context of a single document, i.e., a single web page. The organization of documents within a set is outside the scope of this work.

### 3.1 Focus on Code

> The document format emphasizes high-quality code examples that readers can use to understand a concrete application of software development technologies.

Tutorial authors recognized the importance of good code examples [26] and many of the documentation retrieval and synthesis techniques focus on code examples as the main source of information(e.g., [17, 22, 75]). They capture concrete solutions that anchor the discussion of more abstract concepts or guidelines related to the code. A code example is also a useful starting point for programmers to copy and adapt to accomplish the task they want. For those reasons, programmers commonly choose to first read the code examples of a tutorial, and only refer to surrounding text if they need more information [10]. As a result, documents without code examples, or with code examples that are too simple, are viewed as less helpful by programmers [49, 58]. The benefits of code examples in learning resources are similar, to an extent, to the benefits of video tutorials: they allow the audience to follow along a concrete application of the abstract concepts being discussed [44].

*Implementation.* Casdoc focuses on code by anchoring the hierarchy of annotations in a complete and compilable code example. Other learning resources emphasize code differently. Some tutorials are accompanied by curated sets of standalone examples, intended to integrate the notions described in the tutorial into solutions for more complex scenarios.[6] Other online resources, such as GitHub Gist,[7] are themselves databases of code examples, often with a minimal description of the example's purpose, which can be used on their own or in combination with other documents. Programming "cookbooks" are a more structured version of code example databases.[8] They typically focus on implementation solutions that the reader can adapt to perform common tasks with a technology. Some online learning platforms also guide readers through the implementation of a small program as the main learning activity.[9] However, although such platforms place a higher importance on code, some readers may prefer to see the complete program first, instead of going through each step of the guide. Finally, Oney and Brandt proposed to embed documentation in shareable code fragments, called Codelets [53]. Programmers create Codelets as an HTML document using specific tags to identify links between the code example and its related documentation. Although their idea aims at helping programmers integrate code examples found on the web, rather than as a learning resource directly, Oney and Brandt acknowledge the pedagogical potential of Codelets.

*Limitations.* Code examples alone are often insufficient to describe complex programming tasks. Documents that focus on code should not entirely omit accompanying explanations, which can help the reader adapt the code example to their situation, distinguish important parts of the code from peripheral elements, or learn related concepts. For example, Stack Overflow answers that contain only code often receive downvotes or edit requests to add some explanation of the code [49].

## 3.2 Gradual Reveal

> The document format reveals only a small part of the content at a time, letting the reader understand one fragment before showing the next.

Being overly verbose and containing insufficient information are two common, yet conflicting issues of documentation [2]. Including more information in a document is necessary when the audience is large and varied, as it is often the case for software documents addressed to third-party programmers. However, too much content can have a detrimental effect if it increases the time and effort each reader takes to find the information they need. When readers spend, or estimate that they would spend, too much effort to find the parts of a document they need, they are likely to look for another document [41, 76]. In their idea of the "Minimal Manual", Carroll et al. suggest to "slash the verbiage" [12]: technical writers should reduce the length of manuals by removing redundant and superfluous parts, to avoid readers misusing the documents or missing crucial information. Crowd-sourced documentation platforms in particular, such as Stack Overflow, can accumulate overwhelming content on popular topics. They must find effective ways to emphasize the most important information to readers [77]. Exposing readers to only parts of a document at a time is an alternative solution to this conflict between completeness and verbosity.

*Implementation.* Casdoc gradually reveals its content through annotations in floating popovers or pinned dialogs. The initial view of a Casdoc document shows only the code example, and the reader chooses which annotations to reveal by interacting with their anchors. Collapsible HTML components can also be used for that purpose, allowing readers to choose which information to reveal.[10] Tabbed containers can be useful to include multiple variants of the same content fragment in a document without increasing its bulk. They can show, for example, information to accomplish a task with alternative technologies, allowing readers to select the technology that is relevant to them.[11]

---

[6]E.g., https://www.tutorialspoint.com/javaexamples/index.htm and https://www.w3schools.com/java/java_examples.asp

[7]https://gist.github.com/

[8]E.g., the Python [7] or R [40] cookbooks

[9]E.g., Google Codelabs, https://codelabs.developers.google.com/

[10]E.g., the FAQs document of Amazon API Gateway uses a collapsible element for the answer of each question: https://aws.amazon.com/api-gateway/faqs/

*Limitations.* Revealing information gradually inherently relies on a format that can be modified over time, based on the reader's or timed triggers. Thus, this property may be harder to consistently implement or adapt across software applications and viewing devices. For example, the Casdoc format is designed for mouse interaction in desktop web browsers, and does not support well mobile devices, touch-based interactions, or printing. This single supported context limits the usability of Casdoc documents. Furthermore, dynamic documents are not well suited for long-term archival purposes, unless the viewing technology is archived with them. Thus, it may be useful to produce a static version of dynamic documents as a replacement in situations that do not support user interactions.

### 3.3  Small Fragments

> The document format presents its content as a series of concise fragments that each convey a single self-contained idea.

It is common for programmers to read a document out of sequence [10]: they may look for a specific section related to their needs, skip information that they already know, or go back to an earlier point in the document to find background information about a concept. A set of concise and decoupled fragments supports such reading behaviors. In contrast, documents composed of vaguely bounded and highly dependent fragments force their readers to read larger sections to contextualize and understand the information they seek, which can create a feeling of verbosity. Additionally, identifying clear fragments in a document can facilitate the reuse of the content into other documentation systems (e.g., [17, 33, 75]) or in integrated development environments (e.g., [53, 55]). This reuse scenario expands the value of the document's information beyond its original purpose.

*Implementation.* Casdoc's annotations encourage authors to partition the information into concise fragments that will be presented in small popovers and dialogs. Annotations can link to further supporting explanations in nested annotations, but they should present a complete idea by themselves. Non-interactive formats can also be organized as small fragments. For example, API reference documentation typically contains one fragment per API type or member.[12] Readers are not expected to read the entire API documentation to understand the fragment about a particular element. Question and Answer (Q&A) forums often exhibit this property, as answers are typically created as independent fragments.[13]

*Limitations.* Documents that appear too fragmented can irritate readers [68]. Fragmentation can lead to frustration when readers do not know how to find a fragment of interest, or when they need to gather multiple fragments scattered across a document to answer a query. To prevent this problem, authors should organize the fragments carefully to support an intuitive navigation and place fragments that relate to the same task close to each others in the document's structure. Dividing the content of a document into small fragments can also break its narrative flow or disorient readers that do not know what information they must look for. Thus, this property may be detrimental in some contexts, such as online courses for a homogeneous novice audience.

---

[11]E.g., Android Developer Guides uses this strategy to show equivalent code examples either in Kotlin or Java: https://developer.android.com/guide
[12]E.g., the Java API reference documentation (Javadoc), https://docs.oracle.com/en/java/javase/17/docs/api/index.html
[13]For example, the popular Stack Overflow programming forum, https://stackoverflow.com/questions/

### 3.4 Explicit Hints

> The documentation format includes explicit hints, distinct from textual cues, to help readers understand and navigate the structure of a document.

Navigating within the content of a document is an important aspect of information search [54]. Given the amount of web resources readily available and indexed by search engines, readers have a strong incentive to look for other documents if they do not find the information they seek quickly in the current one. Visual hints of the structure and content of a document reduce the cost of within-document navigation and provide a sense of location and control [66]. They can be an especially important tool to mitigate limitations of other properties, such as fragmentation [68]. This property is more incremental than the previous ones. Multiple types of visual hints can be incrementally added to a format to reveal complementary aspects of the structure.

*Implementation.* Casdoc uses markers to indicate the presence of annotations related to a code element or to a concept. Annotations containing only Javadoc information, however, do not have markers as their presence is predictable. The Adamite annotation tool uses a similar strategy to mark parts of a document that have been annotated by readers [31]. Casdoc also uses indicators such as "pin" icons, breadcrumbs, and highlighting to identify the anchor of a pinned dialog. The XCoS code search approach proposed by Wang et al. presents information related to different aspects of the query (e.g., non functional requirements) to help users navigate the list of results [73]. Although those hints are text-based, they constitute a navigation structure distinct from the main list of results. Existing documents also include recurrent types of alternative hints. For example, a table of contents that remains visible and indicates the current position of a reader as they scroll within a page is useful to convey a sense of location within an overview of the document.[14] API reference documentation uses hyperlinks to relate relevant fragments, such as a function to its parameter and return types.[15] Although hyperlinks are a common feature of many websites, the extremely predictable nature of a link's target in reference documentation makes them an effective mechanism to navigate its structure, as opposed to the arbitrary links in typical documents.

*Limitations.* Structural cues integrated in the main text of a document must be used sparingly. They can bloat the content and dilute its relevant information. Ideally, explicit hints should be clearly separated from the document's content, so that the reader can ignore them once they reached the information they sought. Alternatively, hints that rely only on non textual elements, such as Casdoc's markers, can easily be distinguished from the content. However, the hints' purpose must be intuitive, or they risk confusing the readers. They also limit the accessibility of a document for some readers. For example, readers using a screen reader would be oblivious to Casdoc's markers, and therefore to all of its annotations. Thus, purely visual hints should be complemented with other navigation cues, possibly embedded in the HTML tag attributes.

### 3.5 External Content

> The document format provides a systematic way to integrate information from external sources within its original content without corrupting or misappropriating either source of information.

The extensive prior work on documentation generation and information retrieval (e.g., [39, 55, 67, 75]) constitutes a valuable opportunity to increase the coverage of a document. Formats should be designed to leverage these approaches to reduce the effort of multiple authors documenting similar technologies, similarly to how software development evolved to promote the reuse of software packages (especially when well documented).

---

[14]The Android Developer Guides use such interactive tables of contents, https://developer.android.com/guide/components/fundamentals
[15]For example, Java's API reference, https://docs.oracle.com/en/java/javase/17/docs/api/

*Implementation.* Casdoc automatically integrates API reference documentation as additional annotations. These third-party annotations are identified by special icons and contain a link to the information's source. When the anchor of a third-party annotation overlaps with the anchor of an original annotation, the two are concatenated into a single annotation that clearly distinguishes its two parts. As an alternative, the SISE tool designed by Treude and Robillard integrates information fragments from the Stack Overflow forum at the top of API reference documentation pages [67]. The imported information is presented in a rectangle overlay, and contains links to the original Stack Overflow posts.

*Limitations.* The trustworthiness, authoritativeness, and tone of imported content can differ from the document's original content and vary between external sources. Thus, including content from various sources can create jarring changes for the reader, which can affect the perceived qualities of the original content. Clearly identifying the provenance of external content can mitigate these issues. Attributing proper credit is also an ethical and sometimes legal requirement.

## 4  STUDY DESIGN

We evaluated Casdoc in a field study with undergraduate students enrolled in a software design course. Throughout the course, participants had access to a suite of Casdoc documents that complemented the course material. We analyzed how they navigated within the content of each document to assess the strengths and limitations of Casdoc.

We sought to answer the following research questions:

**RQ1**  Is Casdoc a suitable format for creating learning resources for programmers?

**RQ2.1-RQ2.5**  What is the impact of [property 1-5] on the navigation behavior of the readers of a document? (see Section 3)

The first question assessed whether Casdoc is a valuable addition to the documentation landscape. We answered it by offering participants an alternative baseline format, code examples with static code comments, and comparing the adoption of the two formats. The other questions assessed whether the five key properties described in Section 3 help readers find information within a document. To answer these questions, we instrumented the generated documents to log events when participants interacted with Casdoc's features.

### 4.1  Research Method

Our study falls within the *field experiment* category of Stol and Fitzgerald's framework of research methods [64]. We designed the study to prioritize the ecological validity of the results. We conducted our investigation within a natural setting, i.e., a university course, but manipulated the environment to introduce Casdoc documents. This strategy favors the realism of the setting, while allowing the introduction of new elements, such as the Casdoc format, that do not exist in a purely natural setting.

As participants could freely choose and change the documentation format they used, the study is similar to that of a quasi-experiment with a within-subject design. Beyond the field study, our investigation also integrates some aspects of action research, as we continued to improve the Casdoc format based on our findings, to integrate the new documents as a permanent part of the course material for future students (see Section 5.3).

## 4.2 Participants

The field study took place during two consecutive sections of a third-year undergraduate course on software design with an important programming component. All students enrolled in the course could choose to participate in the study by agreeing to a consent form for the collection of their interaction data.[16]

Students are a subgroup of the target audience for Casdoc, i.e., programmers who are learning software development concepts and the usage of some libraries. Although we do not claim that this sample is representative of all programmers, the participants did not act as proxy for a different population.

Both authors had a teaching role in the first section. This familiarity with the course was crucial to create relevant Casdoc documents. The authors were not involved with the second section.

Given that the investigators had authority over participants of the first section, participants remained completely anonymous throughout both sections of the study. This anonymity was important to avoid an unintentional pressure on students to participate in the study or use a format if they did not feel comfortable. Consequently, we did not collect demographic information to measure specific properties of the sample. However, the population from which the sample is taken is well known. Senior undergraduate students in computer science consists predominantly of young adults with only a few years of programming experience, with a minority having previously done industry internships. Before registering for the software design course, students are expected to be familiar with the programming language of the course, Java, and its standard library, but not necessarily with advanced concepts.

## 4.3 Documents

We used the content of the companion website of the course's textbook [57] to create the corpus of Casdoc documents.[17] The website contains three types of documents, namely lists of exercises, descriptions of their solutions in prose, and 126 code examples: 72 of them implement code described in the textbook (i.e., *chapter code*) and the other 54 implement solutions to the exercises (i.e., *solution code*).

We converted each code example into the Casdoc format and inserted additional explanations as annotations. We did not modify the exercise or solutions, which consist mostly of text. The original code examples sometimes contained code comments. We retained those comments in the converted documents, rather than transforming them into further annotations.

We also converted the annotated code examples to a static baseline format. This format included all new annotations as code comments, but it did not include the API reference documentation for standard Java types to avoid unreasonably large comments. This information is, however, easily accessible via the students' integrated development environments (IDEs) and via the official Java documentation website.

The code examples were available on a public website dedicated to the study. The website showed the first document in the Casdoc format to all participants to encourage them to try the new format and provide a consistent user experience. Once on a document, participants could change between the two formats whenever they wanted to. The website stored the last format used in a browser cookie to open the next document in the same format.

The study website initially contained only the annotated code examples. After observing a low study participation rate during the first section, we added the exercises and solutions, unmodified, to the website. Following this change, the retention rate of participants increased for the second section.

---

[17]https://github.com/prmr/DesignBook

Table 2. Events Collected During the Field Study

| Event | Origin | IDs | Details |
|---|---|---|---|
| Visit any page[a] | server | D | |
| Consent to study | server | P/S | |
| Withdraw consent | server | P | |
| Start new session | server | P/S | |
| Open code example | server | P/S/D | format |
| Change format | server | P/S/D | new format |
| Open/Close annotation | client | P/S/D | annotation ID |
| Interact with marker | client | P/S/D | marker ID |
| Use search widget | client | P/S/D | query; selection(s) |
| Use navigation widgets | client | P/S/D | result |

[a]This event was only collected during the second section of the course.

### 4.4 Data Collection Infrastructure

We instrumented the documents to record traces of the participants' activity. Client-side JavaScript functions created the interaction events and sent them to an HTTP POST endpoint of a dedicated data collection server. Only Casdoc documents had interactive features, and therefore generated interaction events. However, the server also recorded document requests in either format and requests to change the format. These events did not rely on client-side functions. For the second section, we also modified the server to record the total number of requests it received, both to monitor the website's status and to estimate the sampling bias.[18] We report on an analysis of the sampling bias in Section 5.4

This logging mechanism was asynchronous and did not affect the user experience. The usage of Casdoc's features did not require a successful exchange with the website, other than fetching the document with an initial GET request. Thus, problems with the data collection infrastructure would not affect students, unless the entire study website was down.

The study website did not require any form of authentication, or ask for personally identifiable information, to preserve the anonymity of participants. To follow events performed by different participants, we stored two HTTP cookies in each participant's browser, in addition to the format-related cookie. Upon consent, participants received a randomly generated 64-bit integer in a persistent cookie (i.e., the *participant ID*). The website also sent a second random integer in a session cookie (i.e., *session ID*), which was reset every time the browser was reopened.

Table 2 summarizes the types of events we collected. The first six types of events are generated by the website, whereas the last four types are generated by JavaScript functions and sent through the HTTP POST endpoint. For each event, the website stored the type of event and a timestamp, as well as the IDs of the participant (P), session (S), or document (D) and the additional details described in the last column.

This procedure was minimally intrusive to participants. After providing their initial consent, the participants did not see the data collection mechanisms while using the documents. This was a deliberate choice to avoid constant reminders that participants were observed, which could affect their behavior. As a consequence, we did not rely on tools such as pop-up dialogs or surveys to gather the feedback of participants. Thus, although we did not collect insights about the subjective perceptions of participants on Casdoc, the collected data is more likely to represent the true behavior and preferences of participants.

---

[18]We did not observe any unusual access patterns, except for a large number of requests to the website's home page: there were almost six times as many requests to the home page as the number of requests to all code examples combined. These requests could be due to web crawling or server maintenance bots.

Table 3. Summary Statistics of the Collected Data

| Property | Section 1 | Section 2 | Total |
|---|---|---|---|
| Study length (days) | 104 | 102 | 206 |
| All document requests[a] | – | 19 594 | – |
| Code example requests[a] | – | 14 644 | – |
| Enrolled students | 165 | 321 | 486 |
| Participants | 124 | 202 | 326 |
| Sessions | 176 | 541 | 717 |
| Opened code examples | 827 | 6511 | 7338 |
| Logged interactions | 2795 | 15 570 | 18 365 |

[a]including from non participating students

### 4.5 Data Preparation

Table 3 gives an overview of the data we collected. In total, 326 participants generated over 18 000 interaction events. They consulted the 126 code examples a total of 7338 times. We collected more data during the second section of the course, partly due to the larger enrollment and to the changes to the study website made between the two sections. However, both data sets show the same trends (see Section 5.4).

We reassembled the flat list of events into a meaningful structure to analyze our results. The actions of each participant are split into *sessions*, i.e., a period of continuous usage of the website. During a session, a participant *viewed code example documents*. Participants performed different actions on code examples, such as *viewing an annotation* and *using the search widget*. Based on a preliminary inspection of the data, we considered all events performed within 15 minutes of consenting to the study as part of a *learning period*. We excluded the data of all participants who did not interact with the website beyond their learning period.

We initially split sessions using the session ID cookie. However, we found that the cookie was unreliable to track continuous usage. Some participants rarely closed their web browser, creating sessions that spanned many days or weeks. We split such long sessions whenever a participant did not generate any event for two consecutive hours.[19] Within each session, opening a document initiates a new code example view, and all subsequent actions performed on this document are associated with this view. After artificially splitting long sessions, any document that remained opened initiates a new code example view in the second part of the session if the participant performed any action on the document. A single session can contain multiple views of the same document, if it is closed and reopened, and multiple views of different documents can overlap.

We grouped successive events associated with the same Casdoc annotation as a single annotation view action. Each annotation view starts with zero or more hovering events, optionally followed by a pin event, and a final optional unpin event. We grouped together multiple hovering events if they were less than five seconds apart, to account for participants accidentally moving outside the marker and immediately going back to it. To avoid spurious events, a hovering event was generated only if the participant hovered for at least one second over a marker.

As each keystroke in the search widget generated a new event, we grouped all events that incrementally built towards a single search query, as well as subsequent interactions with the search results, as a single search action. Each use of the breadcrumbs and the undo and redo buttons constitutes a separate action.

---

[19]We chose the threshold of two hours based on the distribution of time between two consecutive events with the same session cookie. Nevertheless, as we did not observe a significant drop in the distribution, this threshold is only approximative. We avoid relying too much on sessions in our analysis.

### 4.6 Study Design Trade-Offs

There is an inherent trade-off between the realism of, and control over, the study setting. As the field study favors realism, we could not control when or for how long participants used the documents. Field studies also lack the control of confounding factors that the sterile environment of laboratory experiments provides. Thus, the decision to conduct a field experiment limited the precision of our measurements and generalizability of our results [64], but produced concrete insights that are directly applicable to an existing context. These insights led to the improvement of Casdoc for future students enrolled in the course.

Another early decision point was the choice of the environment in which to conduct the study. We chose to study students enrolled in a university course. Alternative options included looking for programmers outside our organization, such as professional developers, or using remote experimentation platforms such as Prolific.[20] The effort involved in recruiting participants for a long-term study (several months) and creating a realistic environment in which participants would need learning resources was a deciding factor for choosing the university course. A consequence of this decision was the need to mitigate potential pressures on students. We thus designed the data collection to be anonymous and minimally intrusive, which was consistent with the choice of our research method. Recruiting students as participants also narrowed down the sampling frame of our study. Thus, our results are specific to a well-defined subset of Casdoc's target audience, and more experimentation is required to generalize them to more experienced programmers.

The number and choice of document formats to compare was also an important decision. Alternative formats include presenting the additional explanations in a narrative text above, below, or interleaved with the code example, as well as presenting a varying number of explanations in static documents. Offering more formats to participants can help contextualize our observations. However, each format requires a considerable effort to produce, and too many formats can overwhelm participants. We chose to offer one baseline format to have at least one point of comparison for Casdoc. We selected static, commented code examples for the baseline as it is conceptually the closest to Casdoc. We did not vary the content of documents to avoid students missing some relevant information due to their choice of format.

Regarding the collection of events, there is a trade-off between the reliability of the events and the quality of the user experience. Asynchronous client-side functions increase the risk of losing some data, e.g., due to students with an unreliable Internet connection. HTTP cookies can generate inconsistencies in the data, e.g., due to students clearing their cookies during the course. The website's HTTP POST endpoint was also vulnerable to potential attacks from malicious actors, who may try to send fake events. We accepted these risks to improve the user experience, to honor our responsibility to create a suitable learning environment for students in the course. The features of Casdoc were not affected by corrupted cookies or the latency of the study website, as they would have if the content of annotations was retrieved using synchronous requests, for example. To limit and detect the generation of corrupted events, we ensured that key events were generated by the website, such as new document requests. For example, events describing interactions with a document that was never opened would reveal some inconsistencies. We also devised a strategy in which the type of event in the client-side scripts would be encrypted based on the session and participant IDs to make it harder to send undetected fake data to the POST endpoint. We found no inconsistency in the collected data.

## 5 RESULTS

Figure 2 shows a timeline of the participants' activity through the study, and Table 4 presents an overview of the main study artifacts and observations. Although we excluded 122 short-term participants (37.4%, see Section 4.5), we retained interaction data related to 6770 document views by the others. As participants viewed the large majority of documents (96.1%) in the Casdoc format, the data shows clear usage patterns for the different features of Casdoc, but only limited insights into situations that Casdoc does not support well. We present these patterns in Section 5.1, and discuss their implications on documentation formats in Section 5.2. We discuss the aggregated data from both groups together, as participants from the two groups exhibited the same patterns. The study findings informed several improvements to Casdoc, which we detail in Section 5.3. In Section 5.4, we report on an analysis of the differences between the two sections and of the sampling bias.

---

[20] https://www.prolific.co/

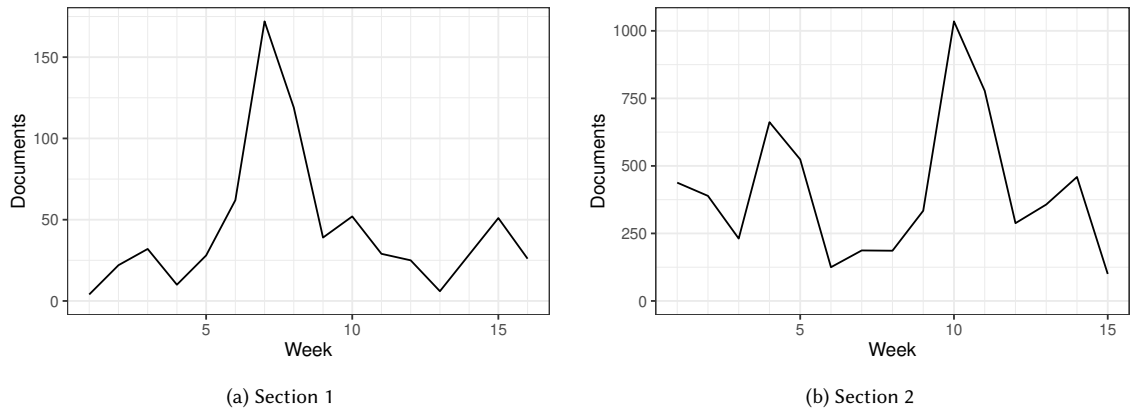

(a) Section 1        (b) Section 2

Fig. 2. Number of Code Examples (Documents) Accessed by Participants During Each Section of the Course

Table 4. Summary Statistics of the Data After Preprocessing

| Property | Section 1 | Section 2 | Total |
|---|---|---|---|
| Participants | 54 | 150 | 204 |
| Sessions | 155 | 1060 | 1215 |
| Unique code examples | 123[a] | 126 | 126 |
| Code example views by all participants | 677 | 6093 | 6770 |
| Code example views in Casdoc format[b] | 670 | 5836 | 6506 |
| Unique original annotations | 417 | 417 | 417 |

[a]For technical reasons, three code examples were not available during the first course section. We fixed this issue for the second section.

[b]Excluding views during which the participant changed format

## 5.1 Casdoc Usage Patterns

Table 5 presents the detailed findings of the field study, grouped according to our research questions.

*RQ1. Viability of Casdoc.* The majority of participants (178 of 204, or 87.3%) used only the Casdoc format throughout the course. Considering that participants had no explicit incentive to use Casdoc rather than the baseline format, this observation strongly suggests that Casdoc is suitable for presenting annotated code examples. Furthermore, among the 26 participants who tried both formats, only six (23%) retained the baseline until the end of the course. Most of the participants who reverted to Casdoc did so within the same session.

Table 5.  Metrics and Results by Research Question

| RQ | Question/Metric | Section 1 | Section 2 | Total |
|---|---|---|---|---|
| 1 | *Viability of Casdoc* | | | |
| | Participants who used only Casdoc | 49 | 129 | 178 |
| | Participants who tried the baseline format | 5 | 21 | 26 |
| | … only during the learning phase | 1 | 5 | 6 |
| | … for only one document after the learning phase | 2 | 8 | 10 |
| | … for only one session (2+ documents) after the learning phase | 1 | 3 | 4 |
| | … for multiple sessions | 1 | 5 | 6 |
| | Participants who changed to the baseline format more than once | 0 | 3 | 3 |
| | Participants who kept the baseline format until the end | 1 | 5 | 6 |

Finding:  *Most participants only used Casdoc. Most of those who tried both formats changed back to Casdoc within the same session.*

| RQ | Question/Metric | Section 1 | Section 2 | Total |
|---|---|---|---|---|
| 2.1 | *Focus on Code Examples* | | | |
| | Server-side code example requests | — | 14 644 | — |
| | … chapter code | — | 8857 | — |
| | … solution code | — | 5787 | — |
| | Server-side exercise statement requests | — | 2539 | — |
| | Server-side solution description requests | — | 2411 | — |
| | Solution code to description requests ratio | — | 2.4 | — |
| | Average number of links to solution code per solution description | — | 3.8 | — |

Finding:  *Participants found value in documents centered around code examples, to support the rest of the course material.*

| RQ | Question/Metric | Section 1 | Section 2 | Total |
|---|---|---|---|---|
| 2.2 | *Reveal Information Gradually* | | | |
| | Annotation views | 356 | 1889 | 2245 |
| | Participants who used annotations | 35 | 115 | 150 |
| | % annotated document views with 1+ annotation view(s) | 18.8% | 15.6% | 15.9% |
| | % markers in the code example interacted with (average, by participant) | 9.3% | 9.1% | 9.1% |
| | % unique original annotations viewed by at least one participant | 23.0% | 60.9% | —* |

Finding:  *Most participants used annotations to find further information about elements of the code examples, but only for a minority of the documents they looked at.*

* We did not aggregate the coverage of unique annotation over the two sections as the documents changed slightly between the sections.

| RQ | Question/Metric | Section 1 | Section 2 | Total |
|---|---|---|---|---|
| 2.3 | *Split Information into Small Fragments* | | | |
| | Annotation views | 356 | 1889 | 2245 |
| | … viewed by only hovering on the anchor | 311 | 1632 | 1943 |
| | … viewed by clicking on the anchor | 43 | 227 | 270 |
| | … viewed without interacting with the anchor | 2 | 30 | 32 |
| | Original annotation viewed from the anchor | 228 | 1385 | 1613 |
| | … with a nested anchor | 41 | 131 | 172 |
| | … with an anchor in the code example | 187 | 1254 | 1441 |

Finding:  *Participants mostly viewed annotations in floating boxes, suggesting that they can grasp the information quickly. Participants viewed nested annotations at a relative rate similar to top level annotations.*

| RQ | Question/Metric | Section 1 | Section 2 | Total |
|---|---|---|---|---|
| 2.4 | *Structure Information with Explicit Hints* | | | |
| | Breadcrumbs used | 0 | 1 | 1 |
| | Undo/redo buttons used | 0 | 1 | 1 |
| | Search queries | 4 | 208 | 212 |
| | … where the participant hovered over the results without selecting one | 0 | 32 | 32 |
| | … where the participant selected at least one result | 2 | 20 | 22 |
| | Participants who used the search bar at least once | 3 | 43 | 46 |
| | Document views with at least one search query | 3 | 159 | 162 |
| | % inline (vs block) markers seen by participants (average) | 57.8% | 61.1% | 60.3% |
| | % inline (vs block) markers interacted with by participants (average) | 85.6% | 86.9% | 86.6% |

Finding:  *Participants did not rely often on secondary navigation aids, suggesting that the markers are effective navigation hints. However, small differences in the visual appearance of markers impacted their effectiveness.*

| RQ | Question/Metric | Section 1 | Section 2 | Total |
|---|---|---|---|---|
| 2.5 | *Support the Integration of External Content* | | | |
| | Unique annotations in all documents | 1565 | 1529 | — |
| | … with only Javadoc content | 1148 | 1112 | — |
| | Annotation views | 356 | 1889 | 2245 |
| | … with only Javadoc content | 128 | 485 | 613 |

Finding:  *API reference documentation augmented code examples with a considerable number of annotations without additional effort. These imported annotations were used by participants, representing a quarter to over a third of all annotation views.*

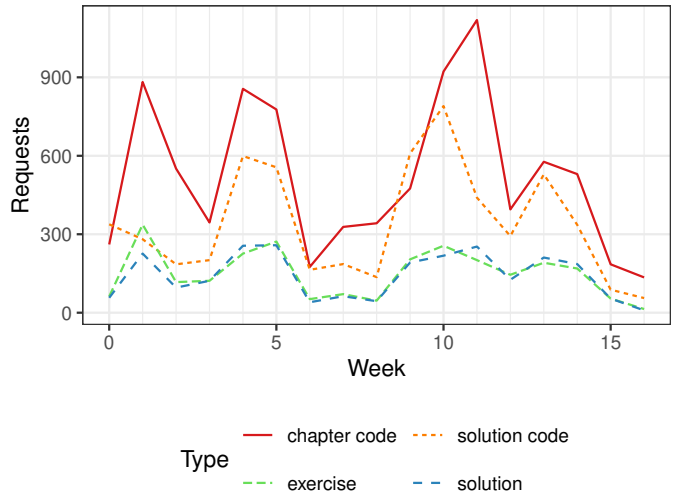

We investigated the documents that triggered a format change and the participants that chose the baseline format over Casdoc to identify scenarios that Casdoc does not support well. However, we found no clear trend in the type of documents (i.e., chapter code or solution code), number of annotations in the documents, or whether the document was among the first ones read by the participants. For example, some participants switched to the baseline on their first document after the learning period, whereas others only tried the baseline after having read over 80 documents already.

*RQ2.1. Focus on Code Example.* Both Casdoc and the baseline format presented a central code example, which was not associated with an interactive feature. Thus, instead of using the participants' interaction data, we compared the number of requests for code examples to requests for other documents to assess the value of code-oriented documents. As the study website did not log requests during the first section, this analysis relies only on the second section of the course.

Figure 3 shows the number of weekly requests for each document type. Students looked at code examples, in particular chapter code, almost three times as often as exercise and solution descriptions. The number of requests fluctuated over time, but the usage of code examples was strongly correlated to the exercise and solution descriptions (Pearson's $r = 0.93$ between the distributions of weekly requests), suggesting a consistent usage of all types of documents.[21] Comparing the solution code and description requests can provide more detailed insights into the usage of text-oriented and code-oriented documents. There were nine solution descriptions (i.e., one document per chapter), each linking to an average of 3.8 solution code examples. Each solution description document presents the complete solution to all exercises, with smaller code fragments to illustrate the main part of a solution. However, we still observed that, for each chapter, participants requested solution code examples 2.4 times more often than solution descriptions on average. This ratio did not vary considerably per chapter, even for the two chapters where the solution descriptions did not include any link to solution code (ratios of 2.1 and 1.7). For each chapter, the ratio of solution code to solution description requests ranged from 0.9 and 1.5 (first two chapters) to 4.1 and 3.0 (chapters four and five, for which the description had the most links). The consistent popularity of code examples demonstrates their value as part of the learning material.

---

[21]Additionally, we did not find evidence that usage of code examples decreased significantly over time (Kendall's $\tau = -0.242$, $p = 0.175$ [34]).

*RQ2.2. Reveal Information Gradually.* Participants did not often look at the content of annotations in a code example. Although 150 participants (73.5%) used annotations at least once, they looked at any annotation in only 15.9% of the code examples they consulted (excluding code examples that did not contain any annotation). Furthermore, considering only annotations with a visible marker in the code example, i.e., annotations that are not nested inside others or Javadoc annotations, participants opened only 9.1% of the annotations they saw. This suggests that annotations, a key feature of Casdoc, was not systematically used by participants.

However, we did not expect, or intend, participants to open most of the annotations they saw. One objective of Casdoc is to be able to contain a large amount of information, including fragments that may only be relevant to a small fraction of the audience, without distracting most readers. Hence, although the low usage of annotations may suggest that readers are more likely to miss information in annotations, it also suggests that the gradual reveal of information is effective to avoid overwhelming readers with information that is less pertinent to them. The annotations viewed by participants varied, as they collectively viewed 23.0% and 60.9% of all original annotations (including nested annotations) during the first and second section, respectively.[22] This further suggests that participants effectively adapted the information they saw to their needs, and that the authoring effort of annotation was not wasted.

Finally, looking at each participant individually, we also observed some differences in their behavior. In particular, some participants used annotations much more than the average. For example, ten participants consulted five or more annotations on at least 10% of the code examples they looked at (excluding code examples with no annotation at all), and four participants interacted with more than half of all the visible markers anchored in the code examples. This observation reinforces the hypothesis that Casdoc can adapt the content of a document to individual readers.

*RQ2.3. Split Information into Small Fragments.* When viewing an annotation using its anchor (instead of using a navigation tool), most of the time (87.8%), participants only viewed the annotation in its floating form, i.e., by hovering over the anchor, rather than in its pinned form. Floating annotations are intended for faster interactions with the content of the document, thus suggesting that concise fragments allow readers to quickly grasp the key information in an annotation or identify that it is not relevant to them.

Fragmentation can also make it hard for readers to collect all the information they need. The ratio of nested to top level annotations views was 0.119, which is comparable, and even slightly higher, than the ratio of code example markers participants interacted with (see RQ2.2).[23] Thus, we reach a similar conclusion as for RQ2.2, that the fragmentation can mitigate distractions to readers, but also increases the probably that a reader will miss useful information.

*RQ2.4. Structure Information with Explicit Hints.* Participants rarely used the secondary navigation aids. We observed only one instance of a participant using the breadcrumbs and the undo and redo buttons. The search bar was used more often, slightly over 200 times, by 46 participants (22.5%) in 162 unique document views (2.5%). In 22 cases (10.4%), the participant pinned at least one of the search results. In an additional 32 cases (15.1%), the participant hovered over the search results to reveal the content of the retrieved annotation, similarly to hovering over an annotation anchor. Hence, the search bar was the most useful of the navigation aids, helping participants find information in 10.4% to 25.5% of cases.[24] This is consistent with Feng et al.'s observation that search widgets help users interact more with interactive visualizations [21]: In a study with 830 participants asked to explore data visualizations, they found that most of the participants who had access to a search bar used it and looked for more diverse information than the control group. However, in our field study, relatively to all annotation views, the search bar was not often used to find information, suggesting that the explicit hints from markers were effective to help participants find relevant annotations.

---

[22]We did not aggregate the coverage of unique annotation seen by all participants over the two sections, as the documents changed slightly during the modifications of the study website.
[23]We consider only original annotations in this comparison, as Javadoc annotations cannot be nested. We also exclude annotation views opened using navigation tools, as they do not discriminate between nested and top level annotations.

We observed an interesting difference, however, in the type of markers that participants interacted with the most. Original annotations with a block anchor in the code example were marked by a gray bracket in the left margin of the code example, whereas those with an inline anchor were marked by a blue underline. Of all markers seen by participants (excluding nested anchors, which are always inlined), 60.3% were blue underlines. Yet, the annotations that participants interacted with disproportionately had inline markers (86.6%, sign test comparing the inline anchors viewed to those interacted with, per participant: $p < 10^{-15}$). We suspect that this difference may be due to the visual aspect of the two markers. Gray brackets have a lower contrast with the document's background than blue underlines, and they are often physically farther from the code element of interest due to the code indentation.

*RQ2.5. Support the Integration of External Content.* Importing the API reference documentation of standard Java types and members more than tripled the number of annotations available to participants, without requiring any effort from the documents' authors. Javadoc annotations also contributed to a notable fraction (27.3%) of the annotation views. The absence of authoring cost and the concrete benefits strongly encourage the development of techniques to properly import external content in documents. We nevertheless observe that participants viewed original annotations more often than Javadoc annotations, despite their lower number. Thus, imported external content should not dilute the quality and relevance of the original content specially authored for a document.

## 5.2 Implications

The field study showed that programmers are willing to try different types of documents, which encourages the design of new, interactive formats. From our observations, we derived several guidelines that documentation creators can consider when formatting documents. These guidelines informed the design of a new version of Casdoc, which we present in Section 5.3.

We formulate the guidelines as lessons we learned from implementing Casdoc and evaluating it in a field study. Although they are derived from empirical evidence, more work is needed to reliably generalize each guideline to various contexts and types of documentation.

*Guideline 1: Involve readers to mitigate verbosity.* A format that reveals information selectively based on reader interaction mitigates the cost of including more content into a document. During the study, readers typically did not inspect the entire content of documents, indicating that extraneous content had little impact on their experience. Thus, it allows document creators to aim for a wider audience while keeping documents approachable for individual readers, instead of investing effort to optimize the balance between coverage and verbosity.

*Guideline 2: Do not require user actions to reveal important information.* A format that requires user interaction for a reader to reach specific information increases the probability that the reader will miss that information. During the study, readers never viewed a majority of the annotations, including annotations that were clearly indicated by a marker on the code example. Readers also viewed nested annotations, which required more complex interactions, considerably less often than top level annotations. Thus, even simple interaction patterns, such as hovering over a specific part of the document, may prevent a reader from finding some key information. This risk is especially important to mitigate for novice readers who may not correctly identify their needs and the important information in a document. To avoid this threat, the document's creator should place important information in a prominent place that does not require extensive interaction to discover.

---

[24]The modest success rate can be explained partly because we counted successive queries separately. Therefore, if a participant uses $N$ queries before finding the information they need, this will be measured as a success rate of $1/N$. This was necessary as it is impossible to reliably infer whether the information sought by a participant changes between successive queries.

*Guideline 3: Carefully assess the impact of aesthetic decisions.* Small differences in the visual presentation of a component of the format can have a high impact on the readers' behaviors. During the study, participants seemed to notice blue underline markers (for inline anchors) significantly more than gray brackets (for block anchors). Although we designed both types of markers to be similarly subtle in a document, this difference possibly caused a bias towards revealing annotations with inline anchors more often than annotations with block anchors. Document creators should carefully review multiple visual options, and choose the option that optimizes the discoverability of the document's content over stylistic preferences.

*Guideline 4: Favor an intuitive structure over navigation aids.* A format with an explicit and intuitive structure with predictable hints, that readers can navigate to find specific information fragments, mitigates the need for secondary navigation aids. During the study, common navigation aids such as undo and redo buttons and breadcrumbs trails were almost never used to open annotations. Even the search bar was relatively rarely used compared to direct navigation features, despite a large fraction of each document's content being initially hidden from the readers. Thus, format designers should prioritize the improvement of a document's structure, to make it more explicit and intuitive, over the addition or improvement of more navigation aids.

*Guideline 5: Consider external content to augment the content of a document.* When done well, supporting the integration of external content can greatly improve the coverage and quality of a document at a minimal cost. During the study, readers benefited from over three times more annotations thanks to the integration of API reference documentation. However, this must be done carefully and with the proper attribution, as the quality, style, and authoritativeness of the imported content can vary, and the external content was not originally designed for the target document. Nevertheless, authors and format designers should consider techniques to share information across documents, given the rich documentation landscape.

### 5.3  Improving the Casdoc Format

Following the field study, we released the annotated code examples, without the JavaScript data collection functions, on a permanent website for students enrolled in future sections of the undergraduate course.[25] We leveraged the study findings to implement a new version of Casdoc that address some of its original shortcomings.[26] Figure 4 shows one of the documents in the revised format.

According to Guideline 3, we revised the visual elements of Casdoc, taking into consideration their impact on readers. Instead of its original arbitrary color scheme, Casdoc now uses a color scheme for code blocks that is similar to the one from Stack Overflow, a popular forum among programmers (see Figure 4a). This color scheme should be more familiar to programmers, mitigating the adoption cost of a new format. To make block anchors stand out more, their markers

---

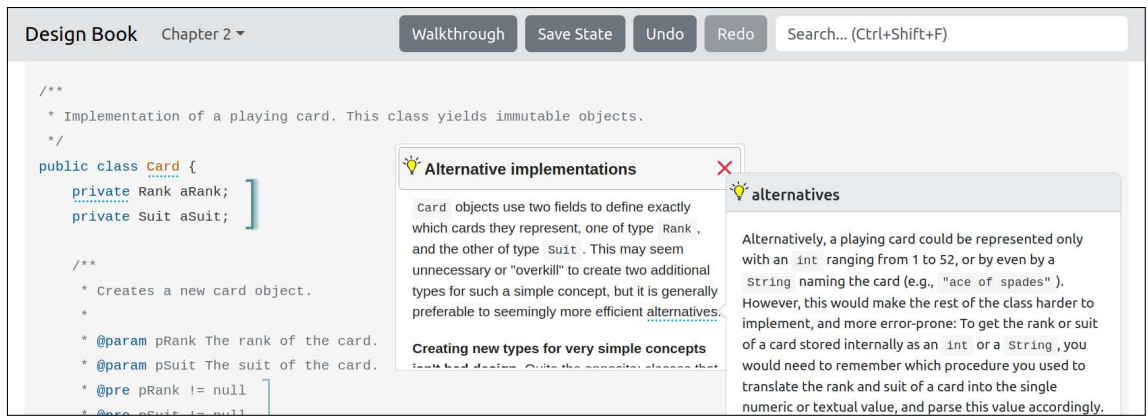

(a) General view of the document with one pinned and one floating annotation

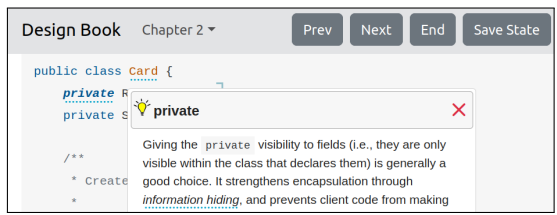

(b) Navigation using the Walkthrough feature

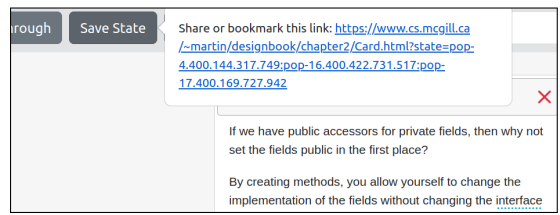

(c) Creation of a link to save and share pinned annotations

Fig. 4. Example of a Document Using the Revised Version of Casdoc

appear at the right of the code and are blue instead of light gray. This increases the contrast of block markers with the background and reduces the distance between the marker and indented code. Finally, anchors of pinned annotations no longer show a "pin" icon next to its marker, as it was disrupting the layout of the code.[27] Instead, inline anchors are shown in bold and italics font and block anchors have their marker in darker blue and with a drop shadow to indicate that the associated annotation is pinned.

We also added three new features to Casdoc. First, authors can now identify a sequence of important annotations when creating a Casdoc document. In addition to their normal behavior, readers can reveal the annotations in such a sequence by clicking on a "Walkthrough" button (see the top of Figure 4a). The walkthrough reveals one annotation at a time, controlled by a pair of "Prev" and "Next" buttons (see Figure 4b). As they move through the sequence, readers can interact with any other annotations. The walkthrough feature was informed by Guideline 2: Although walkthroughs still require some user actions, they allow readers to access important annotations in a standard way across documents, thus reducing the cognitive effort associated to these actions. Consistently with Guideline 4, this feature also provides an additional linear structure to help readers who do not know what information to look for navigate the non linear graph of annotations.

Second, after pinning annotations and laying them out on the page, readers can save the state of the document by clicking on a "Save State" button. Casdoc will generate a custom URL that will reopen the document with all pinned annotations already visible and in the same position (see Figure 4c). Readers can bookmark this link to keep track of annotations they find relevant. The author of a document can also use this feature to manipulate the annotations initially visible to readers. Thus, it can help make important annotations more prominent without requiring any user action (Guideline 2).

---

[27]The icon was also interfering with copy-and-paste behavior, as it would be included in the copied code.

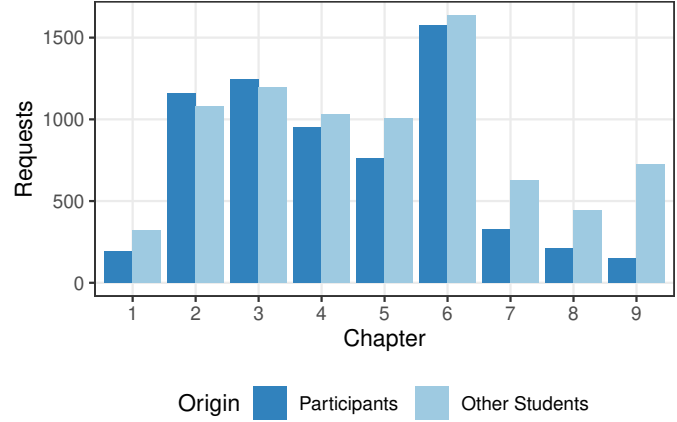

Fig. 5. Code Example Requests by Chapter from Participants and Non-Participants During Section 2

Finally, authors can now store reusable annotations in a database, and insert them in multiple Casdoc documents. In addition to mitigating the need to copy the content of recurrent annotations (e.g., an annotation that describes a common theoretical concept), the database provides a flexible interface to integrate external content into Casdoc documents (Guideline 5). The database stores the exact content of each annotation in an HTML file, associated with another file that contains properties of the annotation, such as its title. Thus, the database can be populated with external content using simple scripts, but the author of a document remains in control of which annotations are added to the document.

### 5.4 Sampling Bias and Differences Between Sections

When recording all document requests received by the study website during the second section of the course, we observed that the majority of requests (55.1%) were not made by participants. As participation in the study was voluntary, there is the risk of a sampling bias in our results. For example, students who are less favorable to trying new technologies may decide to only use the more familiar baseline format and forget to consent to participate in the study.[28] To assess the magnitude of the sampling bias, we investigated whether there was a notable difference between document requests from participants and other students. Figure 5 compares the requests for code examples for each chapter. We observe that the differences are relatively small. A Pearson's $\chi^2$ test confirms that the differences are statistically significant ($p < 10^{-15}$), but the effect size is small (Cramer's $V = 0.18$). The same analysis, but comparing requests by week rather than by chapter, return similar results (Pearson's $\chi^2$ test: $p < 10^{-15}$; Cramer's $V = 0.19$). Thus, although we cannot exclude the effect of a sampling bias on our results, there is no evidence of considerable differences between the sample and the target population.

---

[28]Students had to consent to the study to use the Casdoc format, as the instrumented client-side functions would generate interaction events.

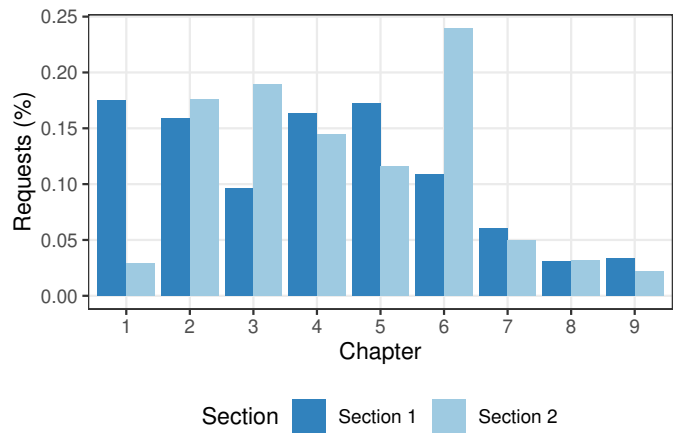

Fig. 6. Code Example Requests by Chapter from Participants of Both Sections

We also compared the requests made by participants from the two sections of the course during which the study took place, to assess how they may differ. We expected some differences, as the two sections were independent of each other. They involved different instructors and students and used a different evaluation scheme and schedule relative to the start of the course. We also released the code examples chapter by chapter during the first section, but all at once during the second section. Figure 6 shows the relative number of requests per chapter for each section.[29] We can indeed observe some differences: The number of requests decreases more consistently during the first section as chapters progress—likely a symptom of the lower retention rate we observed. However, the effect size remains moderate (Pearson's $\chi^2$ test: $p < 10^{-15}$; Cramer's $V = 0.25$), which suggests that the sample of participants did not have a considerable impact on our results.

## 6 RELATED WORK

The importance of documentation in software development and deployment activities motivated a considerable effort to increase the usefulness and quality of documents. During an observational study, Maalej et al. found that developers prefer to look at source code or ask their peers over consulting documentation, but attributed this preference to recurrent documentation issues such as sparsity and a lack of trustworthiness [43]. Knowledge elements such as rationale, intended usage, and real usage scenarios were also often missing from documents. To improve software documentation practices, researchers have studied many aspects of document generation and program comprehension, such as the types of questions asked on public forums [8, 38, 60], the information needs of programmers [9, 19, 27, 62], the overlap between the content of documents and those needs [15], and the types of information provided by different documents [1, 3, 42, 56]. Others have investigated the strategies used by programmers to navigate documents [35–37, 59]. We also discussed additional work related to the generation and usage of documentation in Sections 1 and 3. In comparison, there is much less work on exploring new ways to present information, despite the opportunities created by the migration from printed resources to online documents.

---

[29]We use relative rather than absolute frequencies in this graph for better readability, as the total number of requests was considerably higher during the second section.

Over 30 years ago, Curtis et al. studied different strategies for presenting information about the control flow of small programs [16]. They found that using a constrained language was typically more effective than natural language or ideograms, but the arrangement of the content (i.e., whether it is shown sequentially, hierarchically, or using branches) did not have a considerable impact. Since then, both documentation [69] and software systems have evolved greatly, few researchers reassessed whether their findings still hold in the modern software development landscape. For example, the observations made by Ernst and Robillard in a recent study of architectural documents [20] are consistent with Curtis et al.'s findings, but takes into consideration modern formatting guidelines such as Clements et al.'s Views and Beyond [14]. Many researchers focused instead on new documentation media, in particular video tutorials, as they became popular alternatives to text-based documents [13, 48, 70].

More fundamental work can help build theories to better understand, measure, and predict the quality of software documentation. For example, Sharafi et al. used neuroimaging to identify differences between reading prose and code in brain activation patterns [61]. They also found similarities between programming tasks, such as understanding data structure manipulations, to seemingly unrelated mental exercises such as 3D spatial rotations. In another vein, Hu et al. compared automated metrics used in documentation generation evaluations (e.g., ROUGE, BLEU, METEOR) to human judgment along six quality factors (e.g., naturalness, understandability) [32]. Although they found some correlation between the metrics and human judgment, the automated metrics are not sufficient to completely capture the quality of a document. This line of research is crucial to systematically improve documentation practices, but it is currently insufficient to provide concrete guidance to document authors.

Nevertheless, there has been a resurgence of effort in the human–computer interaction (HCI) field to explore new strategies when presenting text-based online digital documents. These strategies can be used to generate new ideas to improve software-related documents as well. Badam et al. designed Elastic Documents, a system to dynamically link text that describes data to corresponding elements in tables and charts [5]. This technique can help readers make sense of the large and complex data sets underlying data-rich documents. In a laboratory experiment, in which participants were asked to summarize and answer four questions related to two documents, Badam et al. found that Elastic Documents encouraged participants to read more closely the text and figures, whereas participants tended to skim documents more often, even skipping figures, when using a static document viewer. This difference was reflected by a decrease in the quality of the summaries and the accuracy of the answers to the comprehension questions. These results show the potential of adding interactivity to documents.

Similar to Elastic Documents, Masson et al. proposed a system named Charagraph to dynamically generate new visualizations from data presented only in the text of a document [45]. This tool can help readers interact with documents to visualize, compare, and filter the underlying data set of the document. In a laboratory experiment, participants were asked to answer six comprehension questions about six text excerpts from different documents, either using Charagraph or a static document viewer. Although Charagraph did not appear to affect the time required to answer the questions, it significantly decreased the self-reported mental demand, effort, and frustration while increasing the performance of participants, measured with the NASA Task Load Index [25]. It also had a small but significant effect on the accuracy of the participants' answers.

Besides data visualization, Dragicevic et al. discuss the creation of interactive scientific articles that embed multiple data analyses, to allow readers to explore alternatives to what the researchers initially present [18]. Head et al. studied the use of techniques, including interactive ones, to help readers understand mathematical equations in documents [28]. Both of these studies focus on the design of the new format. Among other findings, they provide guidelines for creating, or facilitating the creation of, improved document formats.

Specifically for software documentation, Oney and Brandt designed an approach to create databases of reusable self-documenting code fragments called Codelets [53]. A Codelet combines both the template code to implement a specific programming task and documentation explaining, for example, the purpose of the code fragment or the effect of its parameters. A Codelet can also be interactive, allowing programmers to update the code fragment using a form. In a laboratory experiment, participants using Codelets were able to create a website faster than a control group. This approach is an interesting counterpart to Casdoc: Casdoc documents embed annotations in a code example primarily to create documentation resources. As a result, the documents can provide the initial code for programmers to adapt in their project. In contrast, programmers primarily use Codelets as short fragments to import in their projects, but the documentation ensures that knowledge about the code is not lost during the process.

Blurring the distinction between documentation and data analysis tool even further, Bäuerle et al. proposed Symphony, a framework to create interactive documents that describe machine learning models [6]. A Symphony document allows readers to modify the model and its underlying data set, and see the impact of those changes on associated visualizations. Bäuerle et al. reported on three case studies on the deployment of Symphony, and found that the framework was useful to practitioners as a platform to explore data sets, either for validation, debugging, or as part of learning activities. Symphony also constituted a useful medium to exchange information between different groups of stakeholders in cross-functional teams.

With Casdoc, our goal was to continue this exploration of alternative documentation formats. Instead of a laboratory experiment, we conducted a field study over several months to understand how readers interact with the documents in a natural setting, without the pressure to accomplish specific tasks. This difference in methodology could explain some of the variations in our findings. For example, we found that requiring readers to interact with a document can hide important information, rather than encourage an active reading, as Badam et al. observed [5]. Nevertheless, both our study and prior work show evidence that readers are enthusiastic about dynamic document formats, which motivates future work in this area.

## 7 CONCLUSION

Motivated by the difficulty of creating concise and easily navigable documents that address the needs of a large audience, we designed Casdoc, a novel presentation format for software documents. Casdoc is intended for learning resources that focus on the completion of programming tasks, such as tutorials and usage examples for application programming interfaces. Each document centers around a compilable code example, to which are attached textual annotations that explain its different elements. Readers dynamically reveal and discard annotations by interacting with the elements the annotation describe. As annotations can also be nested within each other, they form a graph that readers can navigate based on specific needs.

We evaluated Casdoc in a field experiment with 326 participants who used over 100 documents during several months. The study focused on the impact of Casdoc's features on the participants' behavior when navigating the content of a document. Although they had access to a baseline format that contained the same information, participants overwhelmingly chose Casdoc as their preferred format. The data collected from their interaction with Casdoc documents allowed us to assess the impact of five documentation format properties on the information that readers consume. We elicited from our observations five guidelines to design new formats, and leveraged those insights to implement an improved version of Casdoc.

Consistently with prior work, our results show that readers appreciate interactive formats for text-based learning resources. However, they also highlight the challenges of creating effective formats. For example, visual elements and the inherent understandability of structural hints should be carefully assessed when designing an interactive format, as they may bias the information that readers are more likely to find. By striving to address these challenges, we aim to increase the overall quality of the documentation landscape for software technologies.

## ACKNOWLEDGMENTS

We are grateful to Zara Horlacher and Emily Shannon for their contribution to the implementation of Casdoc. We also thank the anonymous study participants, our colleagues, and the reviewers and conference attendees for their valuable feedback on different versions of Casdoc. This work was supported by the Natural Sciences and Engineering Research Council of Canada and the Fonds de Recherche du Québec – Nature et technologies.